\documentclass[%
 aip,
 amsmath,amssymb,
 reprint,%
]{revtex4-2}

\usepackage{graphicx}
\usepackage{dcolumn}
\usepackage{bm}

\usepackage[utf8]{inputenc}
\usepackage[T1]{fontenc}
\usepackage{mathptmx}
\usepackage{etoolbox}

\makeatletter
\def\@email#1#2{%
 \endgroup
 \patchcmd{\titleblock@produce}
  {\frontmatter@RRAPformat}
  {\frontmatter@RRAPformat{\produce@RRAP{*#1\href{mailto:#2}{#2}}}\frontmatter@RRAPformat}
  {}{}
}%
\makeatother
\begin{document}

\preprint{AIP/123-QED}

\title{A Low-Temperature Tool for Acousto-electric Detection }
\author{Rohit Mandlik}
 
\author{Sunil Nair}%
\altaffiliation{Author to whom correspondence should be addressed.}
 \email{sunil@iiserpune.ac.in}
\affiliation{ Department of Physics, Indian Institute of Science Education and Research, Dr Homi Bhabha Road, Pune 411008, India 
}%

\date{\today}

\begin{abstract}
\section*{abstract}
We have developed an apparatus for the measurement of the acousto-electric effect (AEE) from room temperature to liquid nitrogen temperatures. The AEE is excited by bulk acoustic waves (BAWs) for which a lithium niobate piezoelectric actuator is employed. A very low background noise level is realized through careful electrical and mechanical design of the setup. The validity of the setup is tested by measuring the AEE in heavily doped P-type and N-type silicon. The opposite sign of the AEE signal was seen for both samples, showing dependence of the signal on the type of charge carrier being dragged by acoustic waves. The signal showed quadratic dependence on the driving voltage, which is consistent with the AEE literature. Further, a temperature dependence of the AEE signal was measured, which exhibits a monotonic increase with a rise in temperature. The ability to probe the AEE in bulk samples at low temperatures makes this setup a promising tool to investigate the AEE and related phenomena in different materials at cryogenic temperatures.

\end{abstract}

\maketitle

\section{\label{sec:level1} Introduction}
The acousto-electric effect (AEE), the generation of a dc voltage along the propagation direction of an acoustic wave in a material with mobile charges, was theoretically predicted by Parmenter in 1953\cite{Parmenter1953} and first experimentally observed in an n-type germanium crystal by Weinreich and White in 1957.\cite{Weinreich_and_white_1957} This effect is a direct consequence of the wave-particle drag mechanism, where acoustic waves drag mobile charges in a material in the direction of propagation, and gives direct insight into the interaction between the acoustic waves and mobile charge carriers\cite{Parmenter1959,Eckstein1964}. The polarity of the AEE signal is determined by majority charge carriers in a material\cite{Weinreich1956}, enabling their identification without the application of an external magnetic field, in contrast to the conventional Hall effect.
Moreover, the acousto-electric interaction also leads to effects such as the anomalous AEE\cite{LCMO_2001,LCMO_2003}, the acoustomagnetic effect\cite{Bismuth1965}, amplification of sound\cite{CdS_PRL_1961_Hutson1961}, the oscillating AEE\cite{Mou2025}, all of which have been experimentally observed. Other effects such as the valley AEE, the valley acoustoelectric Hall effect\cite{Valley_AEE_PRL_2019_Kalameitsev2019}, the acoustogalavnic effect\cite{Dirac_and_weyl_semimetal_2020} and exotic transverse charge current\cite{psedogauge_field_Bhalla2022}, clapping mode-related AC transverse anomalous AEE\cite{clapping_mode_2022} are predicted but remain to be observed. The diverse phenomena associated with the AEE make it a compelling area for further investigations.

The AEE can be measured using two types of acoustic waves: surface acoustic waves (SAWs) and bulk acoustic waves (BAWs). SAWs have been employed in the past to study the AEE in materials such as Si-Lithium Niobate \cite{Si_LN}, GaAs-AlGaAs 2D electron gas\cite{GaAs-AlGaAs_2D_electron_gas_1996}, GaAs-AlGaAs heterostructures\cite{GaAs-AlGaAs_heterostructures_1995}, graphene\cite{gaphene_nanoribons,graphene}, and cadmium sulfide\cite{CdS_1964}. Owing to their confinement near the surface of the piezoelectric substrate and exponential attenuation over a depth of a few micrometers\cite{Mason2014}, SAWs primarily probe near-surface regions and are therefore not ideally suited for investigating the AEE in bulk materials. On the other hand, bulk acoustic waves, which are generated using a piezoelectric actuator and can travel through the bulk of the sample, offer an alternative approach for probing the acousto-electric response of bulk materials. BAW-induced AEE in bulk materials remains relatively less explored, with only a limited number of studies reported till date, including those on Germanium and Bismuth\cite{Bismuth1965}.

Despite its unique ability to directly probe acoustic-wave–carrier interactions, the AEE has been underexplored in the low-temperature regime, where carrier transport, scattering, and relaxation processes exhibit strong temperature dependence. The AEE can also be used to determine the intervalley scattering rate\cite{c_white1959}, band structure of semimetals\cite{Bismuth1965}, dynamic charge modulation\cite{Mou2025}, and the magnetoattenuation of sound\cite{Magnetic_Field_Dependence}. Only a limited number of low-temperature studies of the AEE using SAWs\cite{LCMO_2001,Vandium_oxide_2020,LAO_STO} and BAWs\cite{c_white1959,Bismuth1965,InSb_Kogami1971} have been reported.
In this work, we have utilized BAWs to measure the AEE signal, using a carefully designed low-temperature probe operating in the $80$ to $300\,\mathrm{K}$ temperature range. This probe can be extended to measurements down to liquid helium temperature, while incorporation of non-magnetic components enables future operation in a magnetic field. The utility of the setup is demonstrated by measuring the AEE in heavily doped P-type and N-type silicon.

\section{\label{sec:level2}sample holder design}

 In this arrangement, the piezoelectric actuator, which launches the acoustic waves, is attached to one end of the sample, while an acoustic absorber is placed at the opposite end. The acoustic absorber minimizes the reflection of acoustic waves back into the sample, thereby preventing the resulting backflow of charge carriers that can reduce the measured AEE signal. The mechanical coupling between the piezoelectric transducer and the sample is determined by the mounting pressure. A loosely mounted sample provides weak acoustic coupling, whereas excessive mechanical pressure can damp the piezo vibration and, in extreme cases, may damage the piezo. To meet these requirements, we designed a sample holder in the vertical mounting configuration.

\begin{figure}
    \centering
    \includegraphics[width=1.0\linewidth]{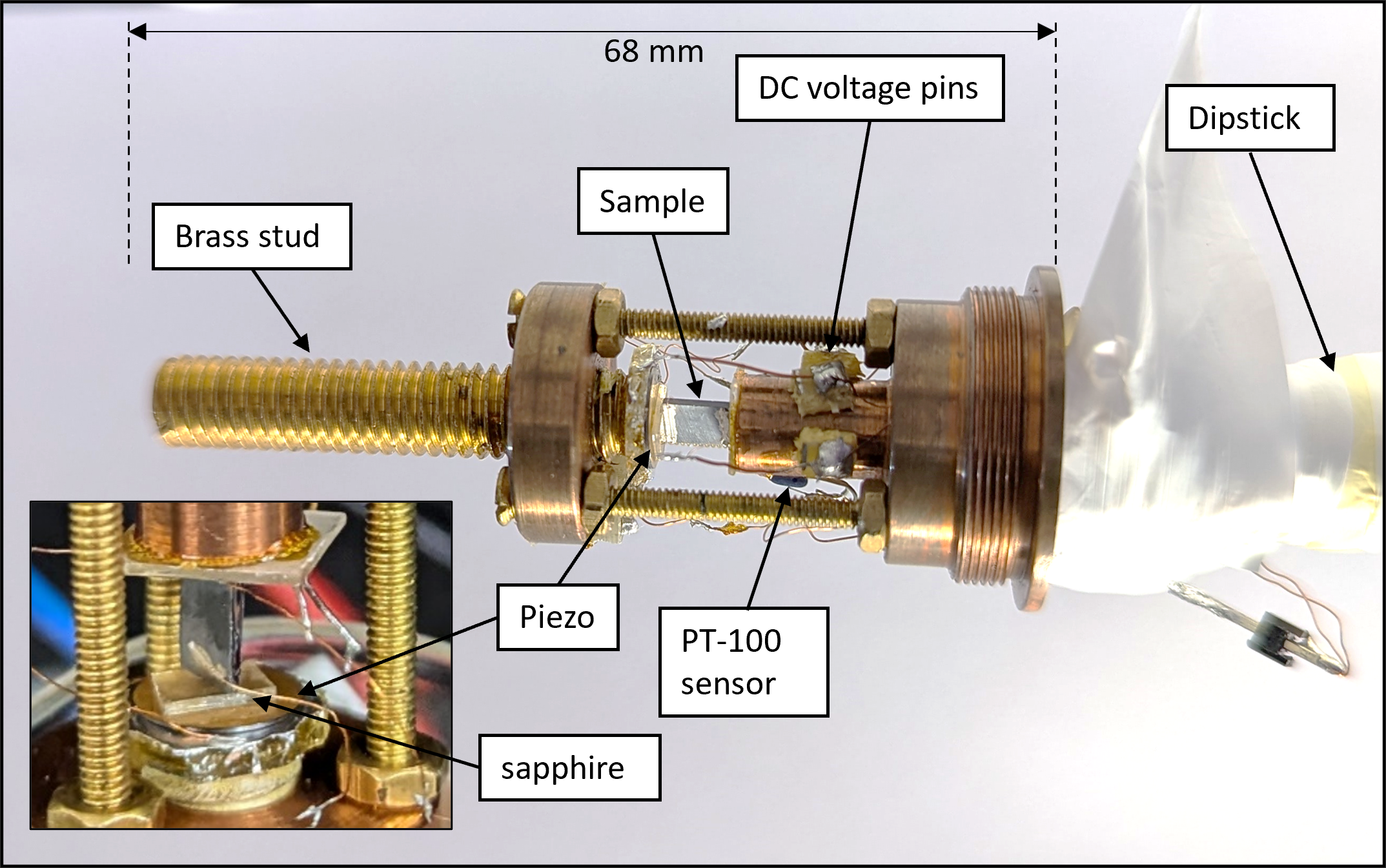}
    \caption{Sample holder with Silicon sample: Different parts of the sample holder are labelled. Threads on the top of the sample holder are for attaching a shield on which the heater is mounted. The sample is mounted on the piezo with the help of a Brass stud. At the bottom of the sample, a piece of sapphire (white) and the piezo (yellow) are shown. On the sample, two connections to the nanovoltmeter are shown. Connections to the function generator, which are used for driving the piezo, are labelled as driving voltage pins.}
    \label{fig:figure1}
\end{figure}

The sample holder was fabricated using copper and brass (Fig.~\ref{fig:figure1}), with a movable M6 brass stud on which the piezo was mounted. While mounting the sample, the brass stud can be moved up and down using a precision screwdriver. Threads on the top part of the sample holder are used to attach a cylindrical shield on which a 25~$\Omega$ manganin wire heater is wound.This shield also ensures no direct contact between the inner part of the sample holder and the external SS cover tube.

The maximum AEE signal is seen at the resonance frequency of the piezo owing to maximum electrical-to-mechanical energy conversion at the resonance.\cite{Weinreich_and_white_1957,26560} Hence, it is imperative to find the resonance frequency of the piezo used to induce the AEE. For this, a Z-cut lithium niobate piezo with a diameter of 6 mm and a thickness of 200 micron was used. Both surfaces of the piezo were coated with a gold-chromium layer. Aluminium foil was affixed to the surface of a glass slide with GE varnish, after which the piezo was adhered to the foil with conductive silver paste. Aluminium foil sandwiched between the piezo and the glass slide serves as a bottom electrode. The large acoustic-impedance mismatch between lithium niobate and glass significantly reduces acoustic transmission through the bottom mounting surface.  Contacts on the top surface of the piezo and on the aluminium foil were made using conducting silver paint. Finally, the piezo mounted on a glass slide was fixed to the brass stud with GE varnish. The sample holder was further dried for 2 hours to ensure proper bonding between the glass slide and brass stud (Fig.~\ref{fig:figure1}).

\begin{figure}[pbth]
    \centering
    \includegraphics[width=1.0\linewidth]{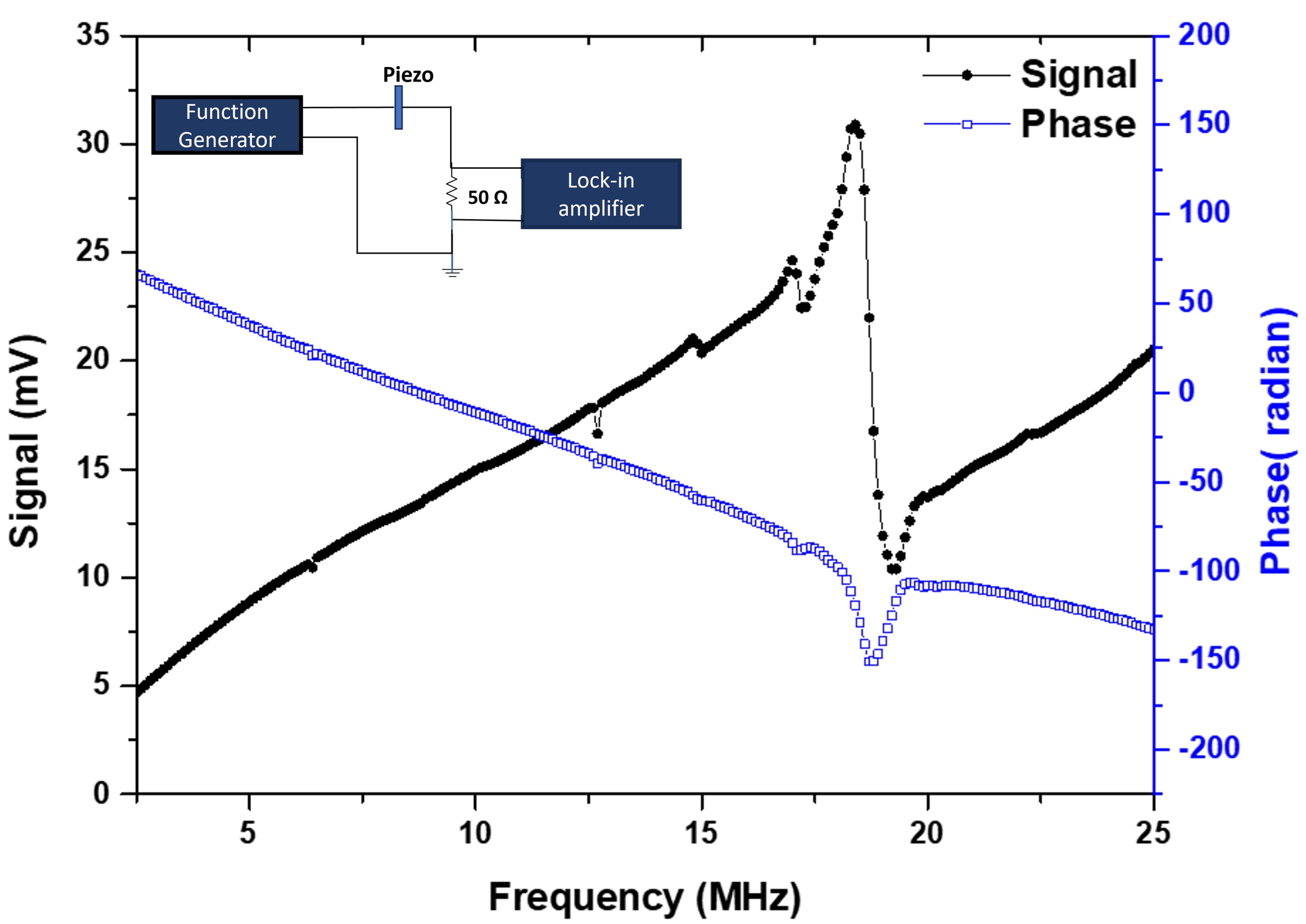}
    \caption{Lithium niobate piezo calibration: the piezo has a resonance at around 18.5 MHz. Near resonance, an abrupt change in phase is seen. The inset shows the circuit used for the piezo calibration.}
    \label{fig:figure2}
\end{figure}

\begin{figure}[pbth]
    \centering
    \includegraphics[width=1.0\linewidth]{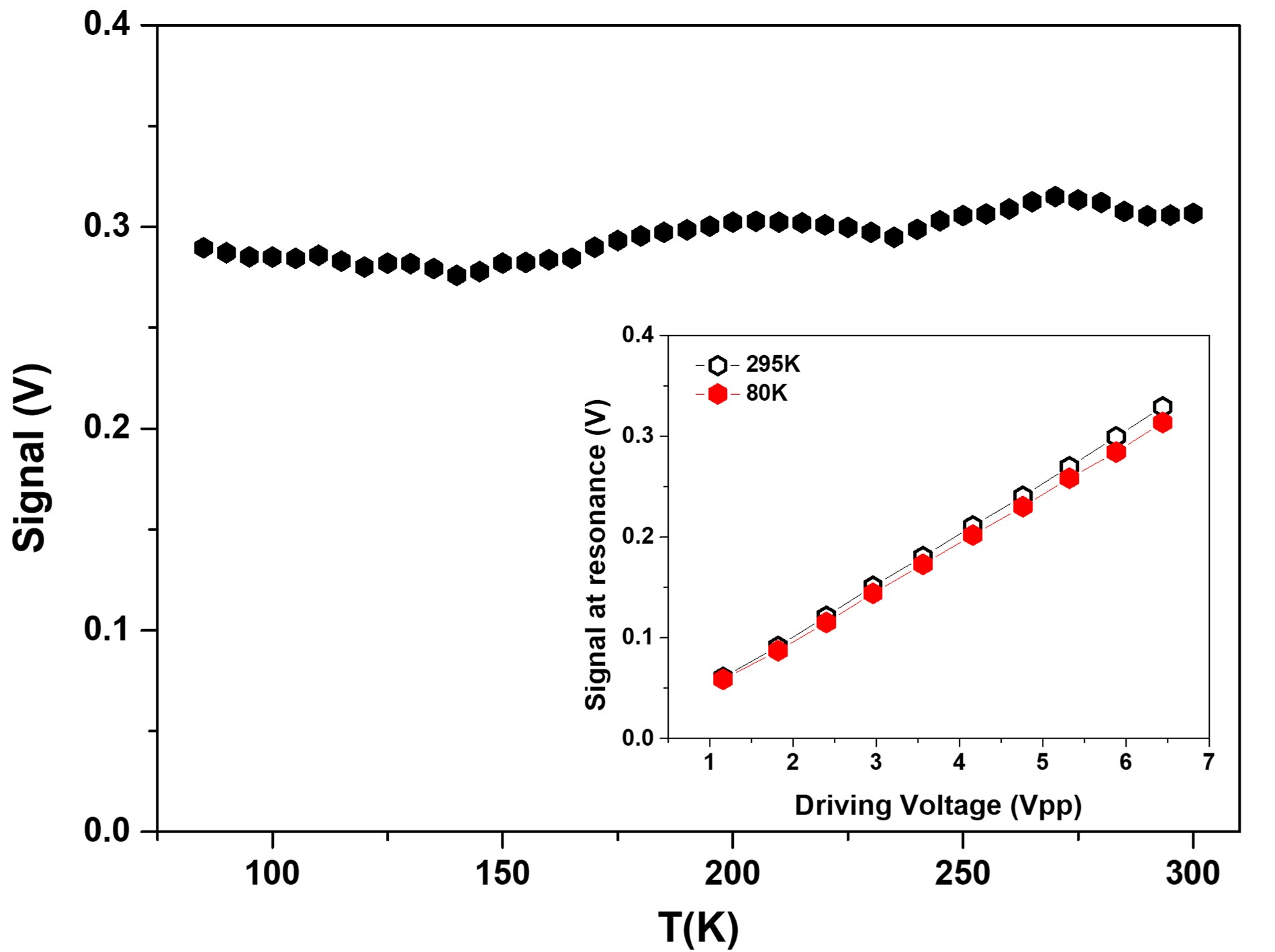}
    \caption{Temperature dependence of piezo response; it shows that piezo response does not change significantly with temperature, and in the inset, dependence of the piezo response on driving voltage is shown, which scales linearly.}
    \label{fig:figure3}
\end{figure}

The lithium niobate piezo was calibrated using the circuit shown in the Fig.~\ref{fig:figure2}. The piezo was driven by a 0.5 Vpp AC signal using a SMG 4022 function generator, and the voltage drop across the resistance while sweeping the frequency from 1 MHz to 25 MHz was measured using a Stanford Research System SR844 RF Lock-In Amplifier. For this, a 50~$\Omega$ RF resistor with an operating frequency range of up to 10 GHz (Part No.:CH0402-50RGFPT) was used. The maximum voltage drop across the resistor and an abrupt change in phase are seen at about 18.5 MHz, which confirms the piezo resonance (Fig.~\ref{fig:figure2}).
Further low-temperature calibration of the piezo from 80 K to 300 K was also done. In order to avoid any extra artifacts and attenuation of the signal, the RF resistor was mounted close to the piezo inside the sample holder at the bottom of the VTI, and a Lake Shore type SC
cryogenic coaxial cable was used to measure the voltage across the resistor. By measuring the temperature dependence of the resistance of the RF resistor, it was confirmed that its resistance does not show any temperature dependence in the 80 K to 300 K range. The piezo was driven by a 5.8 Vpp voltage, and the voltage drop across the RF resistor was measured at different temperatures using a lock-in amplifier (Fig.~\ref{fig:figure3}). This voltage did not show any significant change with temperature. The measured voltage also showed linear dependence on the driving voltage. These observations indicate that the response of the piezo is invariant over the investigated temperature range.

 To maximize the transmission of acoustic waves through the sample, the edges on which the sample is mounted should be flat and parallel to each other. This will also help with proper sandwiching of the sample between the piezo and top copper block. This was achieved by cutting the sample faces with a fine low-speed saw cutter. To measure the voltage drop across the sample, two contacts were made using copper wire and silver paint. For vertical mounting, the sample was held lightly with a tweezer on the piezo, and the bottom screw was gently tightened (Fig.~\ref{fig:figure1}). It was ensured that there was enough tightening force so that the sample would not move easily with mechanical vibrations, while excessive tightening could damage the piezo. After sample mounting, the brass stud was fixed with a GE varnish layer to avoid any movement of the stud during the temperature sweep, which could cause a change in the coupling of the piezo and the sample. To prevent electrical contact between the sample and the piezo, a single crystal sapphire (425~\(\mu\mathrm{m}\)) was added. Another piece of sapphire was placed on the top copper surface to electrically isolate the sample from the sample holder. Owing to the very low acoustic impedance mismatch between sapphire and lithium niobate, most of the acoustic waves were transmitted through sapphire and reach the sample. The top of the sample, held by sapphire on a copper surface, acted as an absorber, ensuring low reflection of acoustic waves.

\section{\label{sec:level2} low temperature instrumentation}

\begin{figure}[b]
    \centering
    \includegraphics[width=1\linewidth]{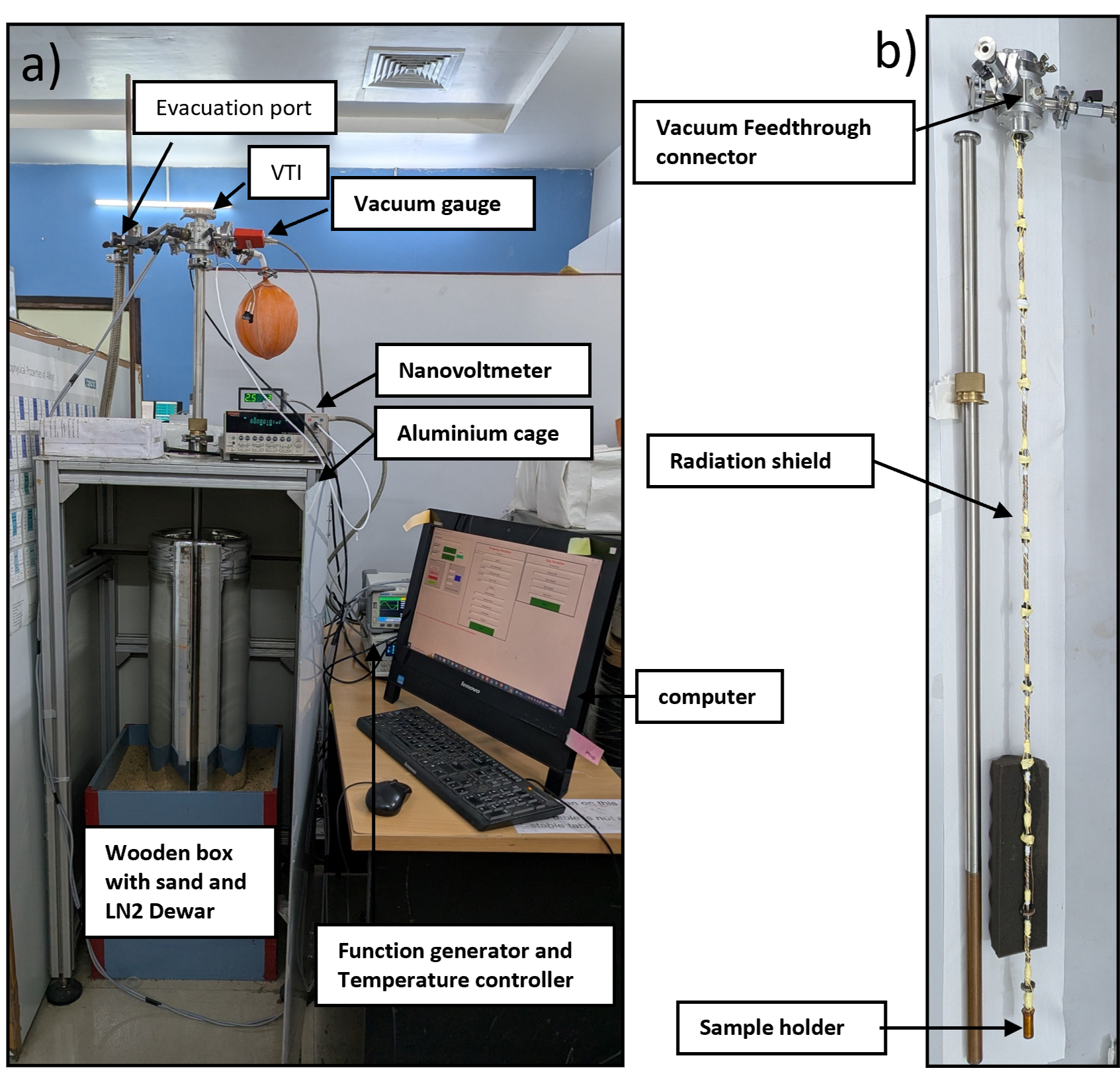}
    \caption{a) Experimental setup: Picture of experimental setup showing VTI, LN2 dewar, and instruments. On the computer screen, an interactive Tkinter-based graphical interface using a Python program can be seen. b) VTI: dipstick with sample holder at bottom and outer SS tube is shown. Yellow patches on the dipstick are wrapped Teflon tape to avoid the movement of wires, which can induce noise.}
    \label{fig:figure4}
\end{figure}

A variable temperature insert (VTI) was designed for cryogenic measurements using a non-magnetic SS 304, ensuring future use in a magnetic field, which includes a dipstick and outer SS cover tube(Fig.~\ref{fig:figure4}b). The dipstick has equally spaced baffles which act as radiation shields. The top part of the dipstick has ports for evacuation and vacuum-compatible feedthrough connectors for AC and DC signals. At the bottom of the dipstick, a hole was made to attach the sample holder using an M3 bolt and nut.
Electrical connections to the PT-100 temperature sensor, heater, and the sample were made using a twisted pair of copper wire. These wires were thermally anchored to the dipstick body.

AC voltage in the MHz frequency range gets attenuated over the length of the wire. To minimize this attenuation, we have used an Amphenol RF coaxial cable of 50 ohm impedance from the SMG 4022 function generator to the BNC feedthrough. Inside a VTI, a Lake Shore type SC coaxial cable was used, which ensures low attenuation at cryogenic temperatures.
Altogether, 3 meters of coaxial cable was used from the function generator to the Piezo, resulting in an attenuation of the driving voltage. This includes losses over the length of wire and reflected power at different joints where impedance mismatch has occurred. The attenuation was measured at the bottom of the VTI using an oscilloscope. For a 10 Vpp applied voltage at 18.5 MHz, 5.88 Vpp could reach the piezo. In addition, we separated the attenuation coming from the cryogenic RF cable using technical specifications of the cable. The attenuation of the signal because of the cable was 0.8 Vpp (0.73 dB), while the attenuation due to impedance mismatch was 3.32 Vpp (3.88 dB). This implies that the majority of the contribution to attenuation of the driving voltage comes from impedance mismatch.

To reduce electrical noise and interference, the bottom part of VTI, including the dewar, was housed inside an aluminum cage, which acted as a Faraday enclosure. Further twisted pairs of copper wire were used for the signal, and the entire setup, including instruments, VTI, and a Faraday enclosure, was grounded using a star connection method. Unwanted movement of wire, which can induce noise, was avoided by anchoring the wires on a dipstick with GE varnish and Teflon tape.

Vibration coming from the ground was isolated by mounting a dewar in a sawdust and sand mixture, while special care was taken to avoid the vibrations coming from the rotary pump. This arrangement together provides the VTI with a mechanical noise-free environment. All measurements were done in a vacuum, and throughout the experiment, the level of liquid nitrogen was maintained above $75\%$ of the total dewar volume.
\begin{figure}[b]
    \centering
    \includegraphics[width=1\linewidth]{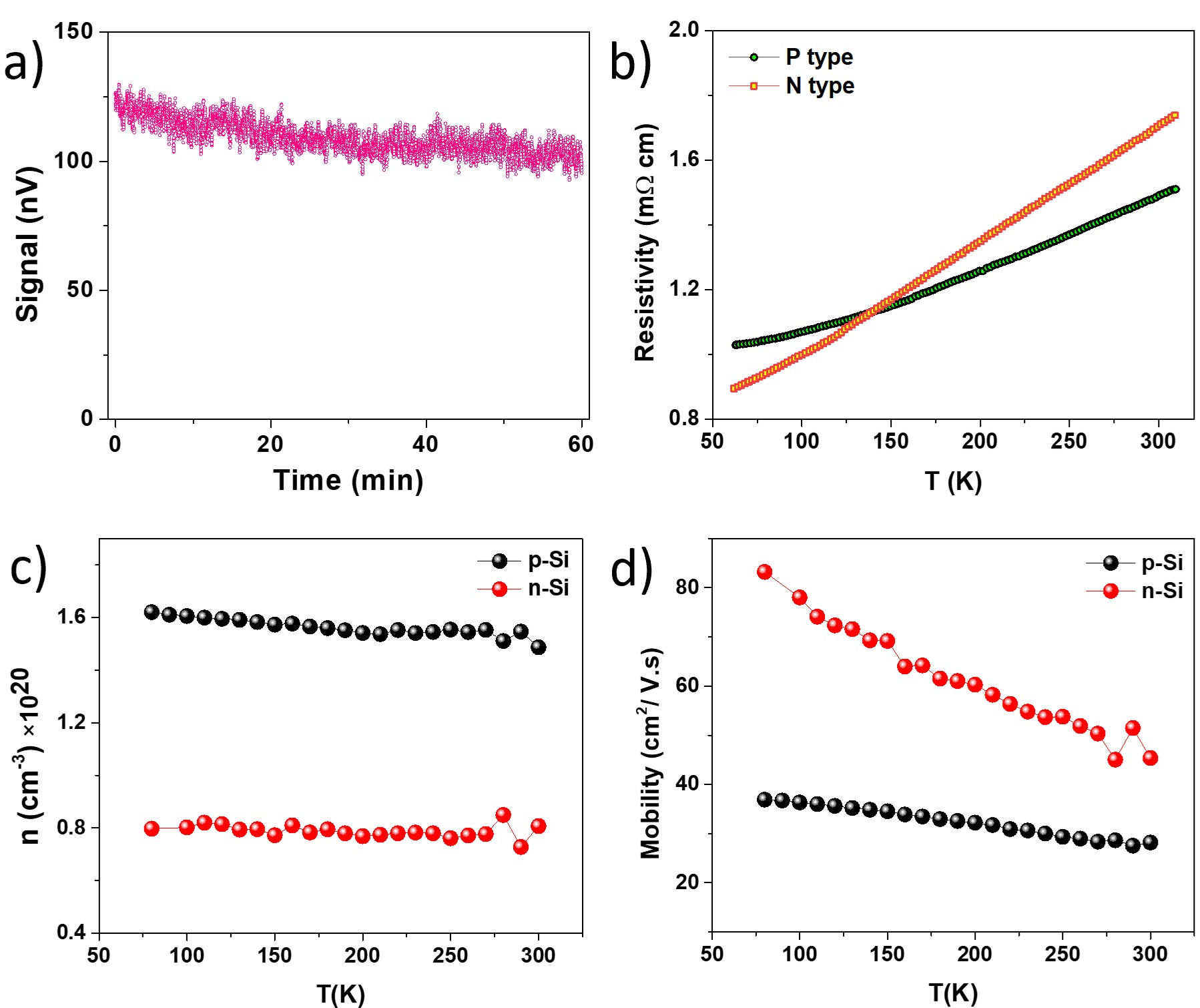}
    \caption{a)Voltage across a P-type silicon sample, when no driving voltage is applied to the piezo at room temperature, shows a stable background over time. b) Low-temperature resistivity of p-Si and n-Si, c) and d) show a temperature dependence of carrier concentration and mobility for holes and electrons for p-Si and n-Si, respectively.}
    \label{fig:figure5}
\end{figure}

\section{Measurement of Acousto-electric Effect}
\label{sec:Measurement_method}
Acousto-electric voltage across the sample was measured by applying the AC driving voltage to the piezo with a function generator, and the voltage drop across the sample was measured using a Keithley Model 2182A Nanovoltmeter. Data were taken in two steps: first, a frequency sweep was measured from 13 MHz to 25 MHz in 0.1 MHz steps, in which the driving voltage (Vpp) was held constant. In the second step, the frequency corresponding to maximum acoustoelectric voltage was detected, and the driving voltage dependence of the signal at this frequency was measured. Data acquisition was automated through a Python program with an interactive Tkinter-based graphical interface, where the code was designed to measure the frequency and driving-voltage dependence of the AEE signal over a specified temperature range.

Using the above setup, we investigated the AEE signal in silicon single crystals of P-type (p-Si) and N-type (n-Si). 
At room temperature, the resistivity of p-Si is $1.5\,\mathrm{m\Omega\,cm}$ while that of n-Si is $1.7\,\mathrm{m\Omega\,cm}$. The temperature-dependent resistivity shows a linear decrease in resistivity with a decrease in temperature(Fig.~\ref{fig:figure5}b). The presence of different types of charge carriers: for p-Si, the majority charge carriers are holes, while for n-Si, the majority charge carriers are electrons, and the absence of any kind of transition in the liquid nitrogen range makes this sample a good choice for the AEE measurement\cite{hole_mobility, Elecrton_mobility}. In these samples, the AEE signal should show opposite polarity, as the sign of the AEE signal is determined by the type of charge carriers in the material. Any structural and magnetic transitions can affect the underlying acoustoelectric interaction by modifying the attenuation coefficient, causing a change in the strength and trend of the AEE signal before and after the transition. The AEE signal in both these samples should show monotonic temperature dependence owing to the absence of such transitions.
\begin{figure}[t]
    \centering
    \includegraphics[width=\linewidth]{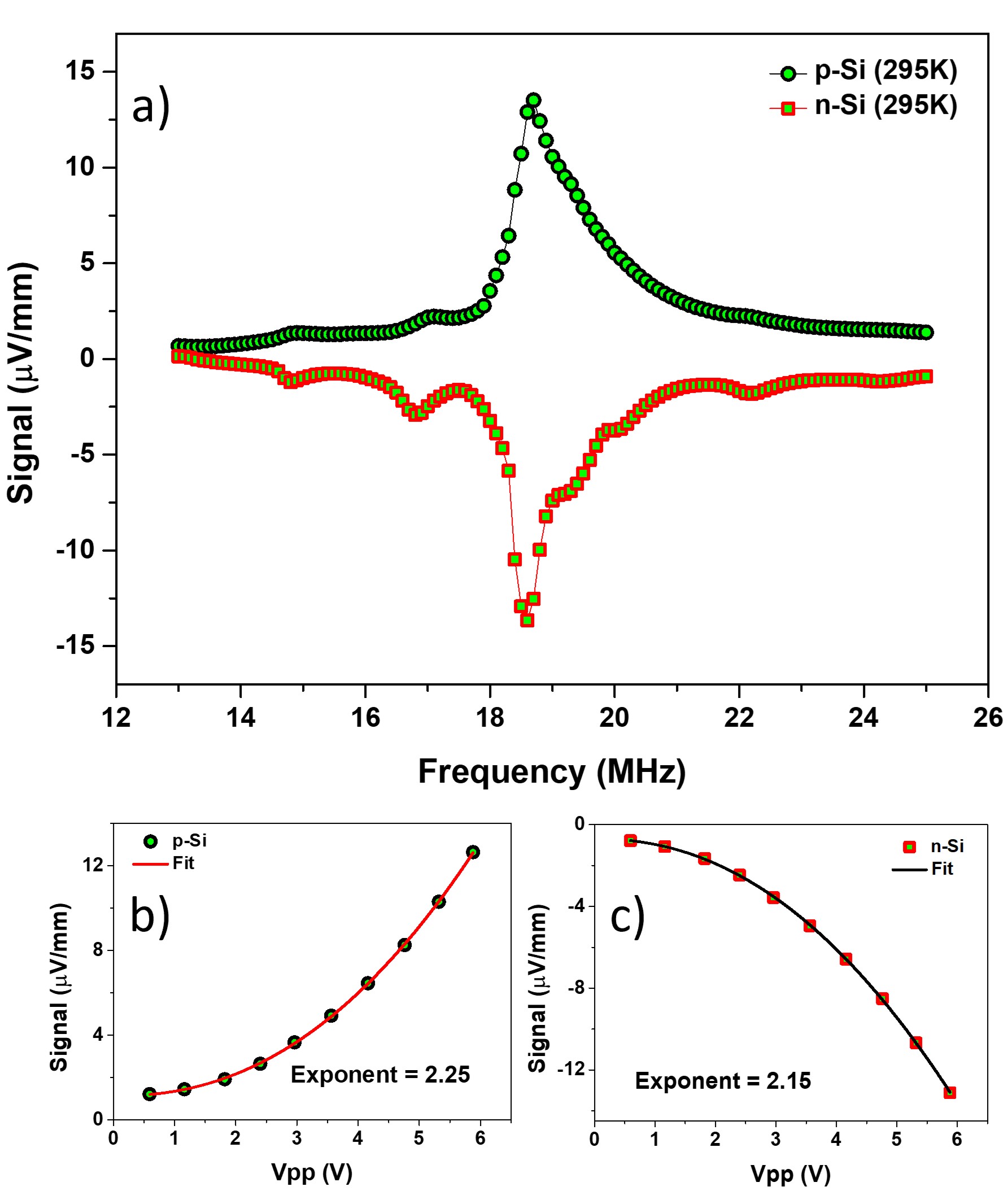}
    \caption{a)Room temperature AEE data in p-Si and n-Si; frequency sweep shows that the peak of the observed signal is at the resonance of the piezo. b) and c) Vpp dependence shows the quadratic nature of the signal. Signal is proportional to $V_{\mathrm{pp}}^{2}$, which implies direct dependence on acoustic energy density}
    \label{fig:figure6}
\end{figure}

For this measurement, the sample was prepared by cutting a 500~$\mu$m-thick silicon substrate into the size of $4\,\mathrm{mm} \times 4\,\mathrm{mm}$ squares. Fig.~\ref{fig:figure5}a shows a stable background signal across the p-Si when the piezo was not excited by driving voltage. The background signal has oscillations in a 10-15 nV range, which suggests that any signal above this should be detectable.
Fig.~\ref{fig:figure6}a shows the AEE signal for both samples measured during the frequency sweep. The AEE signal shows maxima at the resonance position of the piezo. Also, the signal is positive for p-Si and negative for n-Si. The opposite polarities of the signal in the two samples are a direct consequence of the type of charge carrier dragged by acoustic waves, which confirms that holes and electrons are the majority carriers for p-Si and n-Si, respectively.
In Fig.~\ref{fig:figure6} b) and c), the variation of signal with the driving voltage is plotted, which shows that the signal is proportional to the square of the driving voltage ($V_{\mathrm{pp}}^{2}$). The acoustic power density is directly proportional to $V_{\mathrm{pp}}^{2}$, implying that the observed signal exhibits direct dependence on acoustic power density\cite{Bismuth1965}. The occurrence of a signal at the piezo's resonance, linear scaling with acoustic power density, and dependence of the sign of the signal on the type of charge carrier are the key features of the AEE, all of which are observed here.\cite{c_white1959,Bismuth1965,Parmenter1959,Weinreich_and_white_1957,Spector_1963,Weinreich1956} 

\begin{figure}[b]
    \centering
    \includegraphics[width=\linewidth]{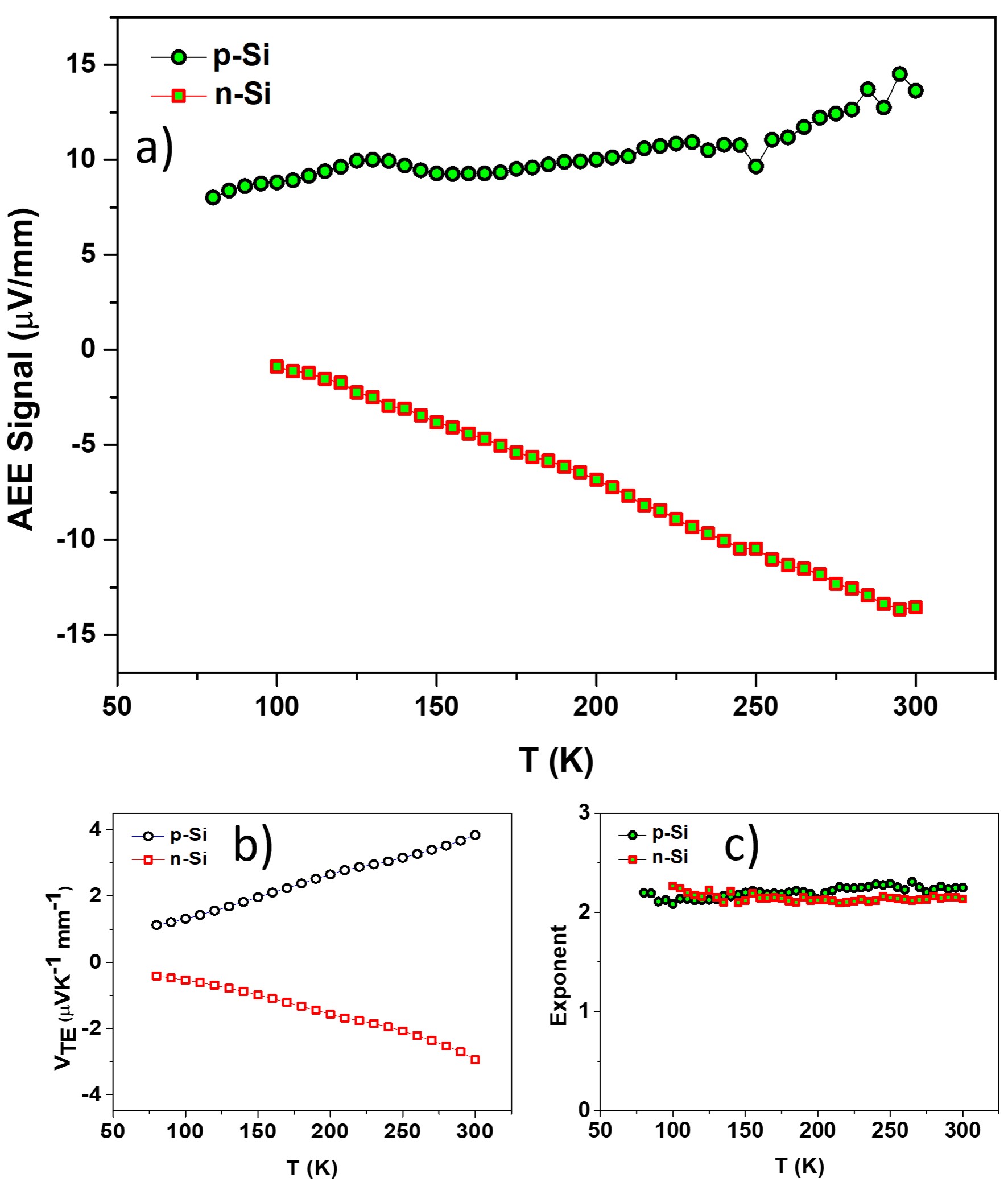}
    \caption{a)Temperature dependence of the AEE signal for P-type and N-type silicon samples is measured; b)Thermoelectric voltage across p-Si and n-Si are measured for $\Delta T = 10\,\mathrm{K}$.; c)Temperature dependence of Exponent c ($V_{\mathrm{pp}}^{c}$), which shows a quadratic dependence on driving voltage throughout the temperature range for both samples. }
    \label{fig:figure7}
\end{figure}

We further investigated the AEE at low temperatures in both samples. The maximum value of the AEE signal at resonance is plotted against the temperature in Fig.~\ref{fig:figure7}a. From low temperature to room temperature, the signal increases steadily. Also, the temperature dependence of the AEE signal shows a quadratic dependence on driving voltage throughout the temperature range, substantiating the direct dependence on acoustic energy density (Fig.~\ref{fig:figure7}c). Both observations are consistent with the assumption that the AEE signal should not show any abrupt change owing to the absence of any kind of temperature-dependent structural or magnetic transition in both p-Si and n-Si. 
With a model of momentum and energy transfer from acoustic waves to charge carriers, Weinreich gave a relation for the acousto-electric field ($E_{\mathrm{ae}}$) and the acousto-electric current ($j_{\mathrm{ae}}$) \cite{Weinreich1957_02,Magnetic_Field_Dependence}, 
\begin{align}
eE_{\mathrm{ae}} &= \frac{\alpha W}{s n_0},
\label{eq:weinreich1}\\
j_{\mathrm{ae}} &= \frac{\alpha \mu W}{s}.
\label{eq:weinreich2}
\end{align}

where $\alpha$ is the attenuation coefficient, $\mu$ is the carrier mobility, $W$ is the acoustic energy density, and s is the sound velocity. From the piezo calibration, acoustic energy density is almost constant in the measured temperature region. The resistivity of both samples decreases with temperature  (Fig.~\ref{fig:figure5}b), indicating a decrease in carrier concentration or/and an increase in mobility. Hall measurements on both samples show that the charge carrier concentration remains nearly constant throughout the temperature range while the mobility increases as the temperature is lowered(Figs.~\ref{fig:figure5}c and~\ref{fig:figure5}d). A possible reason for a decrease in the AEE signal when lowering the temperature could be a decrease in the attenuation coefficient. This will directly affect the energy transfer from acoustic waves to charge carriers. In addition, there will be a contribution from the screening of the electric field associated with acoustic waves by charge carriers. This contribution is particularly important when carrier concentration is high, which may cause a further decrease in the acousto-electric interaction\cite{Kuzmany1974}. The increase in carrier mobility upon cooling may lead to an enhancement in the screening effect, causing a reduction in the AEE signal. The mobility enhancement is more pronounced in n-Si than in p-Si(Fig.~\ref {fig:figure5}d),  which might be causing a stronger screening effect, resulting in a sharper decrease in the AEE signal for n-Si.

Due to the maximum vibration of the piezo at resonance, there are chances of a rise in its temperature. This can induce a parasitic thermoelectric voltage across the sample. It is necessary to estimate the contribution of this thermoelectric signal in the measured AEE signal. By applying a 10 K temperature difference across the sample, we measured the thermoelectric voltage in both samples ( Fig.~\ref{fig:figure7}b). We also measured the temperature difference across the sample when the sample is mounted on a piezo at resonance using an E type differential thermocouple. For p-Si at 300 K and 80 K, the temperature difference across the sample was 0.12 K and 0.05 K, which gives the thermoelectric contribution of about 0.5 $\mu\mathrm{V\,mm^{-1}}$ and 0.05 $\mu\mathrm{V\,mm^{-1}}$, respectively. This voltage is less than 4 $\%$ of the total measured AEE voltage at room temperature. This indicates that a possible thermoelectric voltage contribution is only a minor fraction of the measured AEE Signal. Also, due to the very low Seebeck coefficient of metals, this spurious contribution is expected to be even lower in metals.

\section{Conclusions}
We have successfully developed a fully automated setup capable of measuring the acousto-electric effect (AEE) in a temperature range of 80- 300 K, using bulk acoustic waves (BAWs). The use of BAWs gives this setup the advantage of probing bulk samples, which is otherwise not possible with SAW devices. The setup exhibits a very low background noise level owing to careful electrical isolation and effective suppression of mechanical vibrations. Using this setup, we measured the AEE in P-type and N-type silicon, where the sign of the AEE signal showed dependence on the type of majority charge carrier, while the signal magnitude showed quadratic dependence on driving voltage, showing consistency with classic attributes of the AEE. Further, the observed AEE signal showed a monotonic dependence on temperature and a close correlation with the temperature dependence of mobility. The AEE has been theoretically predicted in many systems such as Dirac/Weyl semimetals\cite{Dirac_and_weyl_semimetal_2020}, and chiral superconductors\cite{clapping_mode_2022}, where this setup can play a significant role.

\section{ACKNOWLEDGMENTS}
R.M. acknowledges CSIR for the financial support provided through a  Senior Research Fellowship 09/0936(11653)/2021-EMR-I. S.N acknowledges DST India for support through Grant No. CRG/2022/003316.

\nocite{*}
\bibliography{bib}

\end{document}